\documentclass[aps,prd,preprint,preprintnumbers,amsmath,amssymb,groupedaddress,nofootinbib,epic]{revtex4}

\usepackage[]{graphicx}
\usepackage[]{epic}
\usepackage[]{eepic}
\usepackage{amscd}
\usepackage[all,cmtip]{xy}
\usepackage{mathrsfs}

\usepackage{simplewick}
\usepackage{changepage}
\usepackage{bm}

\newcounter{RomanNumber}
\newcommand{\MyRoman}[1]{\setcounter{RomanNumber}{#1}\Roman{RomanNumber}}

\newcommand{\bea}{\begin{eqnarray}}
\newcommand{\eea}{\end{eqnarray}}
\newcommand{\beq}{\begin{equation}}
\newcommand{\eeq}{\end{equation}}

\newcommand{\De}{\Delta}
\newcommand{\Db}{\bar D}

\def\<{\langle}
\def\>{\rangle}

\def\nn{\nonumber}

\def\cO {{\cal O}}

\begin{document}

\preprint{}

\title{Bootstrapping Veneziano Amplitude of Vasiliev Theory and $3D$ Bosonization}

\author{ \vspace{8mm}Zhijin Li}
\email{lizhijin18@gmail.com}
\vspace{0.1cm}
\affiliation{
${}$
Centro de F\'\i sica do Porto,
Departamento de F\'\i sicae Astronomia,
Faculdade de Ci\^encias da Universidade do Porto,
Rua do Campo Alegre 687,
4169--007 Porto, Portugal \vspace{2cm}}

\begin{abstract}
\vspace{3mm}
  Three-dimensional conformal field theories (CFTs) with slightly broken higher spin symmetry provide an interesting laboratory to study general properties of CFTs and their roles in the AdS/CFT correspondence. In this work we compute the four-point functions at arbitrary 't Hooft coupling $\lambda$ in the CFTs with slightly broken higher spin symmetry. We use a bootstrap approach based on the approximate higher spin Ward identity. We show that the bootstrap equation is separated into two parts with opposite parity charges, and it leads to a recursion relation for the $\lambda$ expansions of the correlation functions. The $\lambda$ expansions terminate at order $\lambda^2$ and the solutions are exact in $\lambda$. Our work generalizes the approach proposed by Maldacena and Zhiboedov to four-point correlators, and it amounts to an on-shell study for the $3D$ Chern-Simons vector models and the Vasiliev theory in $AdS_4$. Besides, we show that the same results can also be obtained rather simply from bosonization duality of $3D$ Chern-Simons vector models. The odd term at order $O(\lambda)$ in the spinning four-point function  relates to the free boson correlator through a Legendre transformation.
  This provides new evidence on the $3D$ bosonization duality at the spinning four-point function level. We expect this work can be generalized to a complete classification of general four-point functions of single trace currents.

\end{abstract}

\maketitle

\newpage

\tableofcontents

\newpage
\section{Introduction}
One of the simplest examples of AdS/CFT correspondence \cite{Maldacena:1997re,Witten:1998qj,Gubser:1998bc} is provided by the higher spin gravity theory in $AdS_4$ \cite{Vasiliev:1990en, Vasiliev:1999ba, Vasiliev:2003ev, Vasiliev:2012vf} and three-dimensional CFTs with slightly broken higher spin symmetry \cite{Sezgin:2002rt,Klebanov:2002ja,Sezgin:2003pt, Giombi:2009wh,Giombi:2010vg,Giombi:2011ya}. The higher spin symmetry is generated by an infinite set of conserved currents and it leads to strong constraints on the theory. With exact higher spin symmetry, the theory is fixed to be free \cite{Maldacena:2011jn, Boulanger:2013zza, Alba:2013yda, Alba:2015upa}. Given the higher spin symmetry is slightly broken in the sense that the anomalous dimensions of higher spin currents are suppressed by $1/N$ in the large $N$ limit, the conformal correlators are governed by the pseudo-charge conservation identity, or approximate higher spin Ward identity. By solving these equations, the three-point functions of single trace operators can be completely classified into two families \cite{Maldacena:2012sf}: the so called quasi-boson theory with twist $1$ single trace scalar operator and the quasi-fermion theory with twist $2$ single trace scalar operator. The analysis in \cite{Maldacena:2011jn, Maldacena:2012sf} only employs the general consistency conditions of CFTs and the $3D$ higher spin algebra, while does not depend on the details of the microscopic theories, therefore it provides a bootstrap approach to study $3D$ CFTs with (slightly broken) higher spin symmetry.

A class of $3D$ CFTs with slightly broken higher spin symmetry is given by $U(N)_k$ Chern-Simons (CS) theories coupled to a fundamental boson or fermion \cite{Giombi:2011kc, Aharony:2011jz, Aharony:2012nh, GurAri:2012is}.
The theories are solvable in the planar limit ($N,k\rightarrow\infty$ with fixed 't Hooft coupling $\lambda=N/k$) due to the slightly broken higher spin symmetry. In the large $N$ limit, both the CS-scalar  theory and CS-critical fermion theory contain twist $1$ single trace scalar operator and are of the type ``quasi boson" theory according to the classification in \cite{Maldacena:2012sf}. For general $\lambda$,
The two theories share the same three-point functions of single trace currents, up to certain mapping of the parameters and are dual to the same type of Vasiliev theory in $AdS_4$. The CS-fermion and CS-critical boson theories can be obtained from previous two theories through a Legendre transformation. This is the celebrated $3D$ bosonization. At three-point level, this duality has been shown from the bootstrap approach in \cite{Maldacena:2012sf} and explicit field theory computations in \cite{Aharony:2012nh, GurAri:2012is, Giombi:2016zwa, Giombi:2017rhm}. Details on the mapping of the parameters are given in \cite{Aharony:2015mjs} and this duality is conjectured to be true even for finite $N$.
At finite $N$, the CS-boson theory has a beta function for the $J_0^3$ deformation at the order $1/N$. The fate of the IR fixed point
perturbed by this deformation has been studied in \cite{Aharony:2018pjn}.

It is expected that the bootstrap approach \cite{Maldacena:2011jn, Maldacena:2012sf} can also be used to solve conformal four-point functions. The four-point correlators contain information on both single and double trace operators appearing in the OPE. A classification of the four-point functions
will significantly improve our understanding on the dynamics of slightly broken higher spin symmetry.
However, comparing with the three-point functions, the slightly broken higher spin symmetry turns out to be much more complicated for four-point functions. Actually the bootstrap equation for the four-point functions is given by an integro-differential equation with undetermined functions on both sides.
The difference comes from the fact that the three-point functions have already been fixed by the conformal symmetry up to certain OPE coefficients, therefore  the approximate higher spin Ward identities for three-point correlators are essentially a set of algebraic equations of these coefficients; while the four-point correlators contain functions of the conformal invariant cross ratios, which are generically unknown besides certain constraints from permutation symmetry and current conservation laws. The main object of this work is to find a perturbative approach to solve the integro-differential equation.

Alternatively, one may compute the conformal four-point functions using Feynman diagrams from the explicit field theory constructions of the CFTs with slightly broken higher spin symmetry, i.e., the CS-vector models.
In \cite{Bedhotiya:2015uga, Turiaci:2018dht, Yacoby:2018yvy} the scalar four-point correlators in quasi-fermion and quasi-boson theories have been computed  in the planar limit. The computations summed all the Feynman diagrams in the collinear kinematic regime in which the momenta of four external operators are aligned along a particular direction, while the extra two components vanish.
In the quasi-fermion theory, the scalar four-point function is given by free theory up to an overall factor, and the quasi-boson scalar four-point function is a Legendre transformation of the one in quasi-fermion theory, which provides evidence on the $3D$ bosonization at four-point level. In \cite{Turiaci:2018dht}, the authors also employed the analytical bootstrap technique, including the crossing symmetry and the inversion formula \cite{Caron-Huot:2017vep, Simmons-Duffin:2017nub},\footnote{A simple derivation of the inversion formula relating the OPE coefficients to the double discontinuity of the correlation function is provided in \cite{Kravchuk:2018htv}. This relation can also be seen from large spin expansion \cite{Alday:2016njk}. } to fix the scalar four-point correlator based on the known three-point functions up to three truncated solutions to the crossing equation.  The three terms correspond to the contact interactions in AdS and account the non-analyticity of the OPE data in spin. To fix the four-point function completely, the authors resorted to the explicit field theory computations mentioned before. Such result is also obtained in \cite{Aharony:2018npf}, besides, the authors acquired part of the CFT data at order $O(1/N^2)$ from crossing symmetry.
The method based on Feynman diagrams becomes difficult beyond the collinear kinematic regime. For the analytical bootstrap approach, as argued in \cite{Turiaci:2018dht}, if the OPE data has maximal analyticity in spin, then the truncated solutions corresponding the contact Witten diagrams in AdS always disappears, while the AdS exchange diagrams can be fixed from known three-point functions. Nevertheless, the maximal analyticity in spin for general four-point correlators remains an open problem.

In this work, we explore the constraints on the four-point correlators from slightly broken higher spin symmetry, following the approach developed in \cite{Maldacena:2011jn, Maldacena:2012sf}. We will study the quasi-fermion approximate higher spin Ward identity with correlators  $\langle J_0J_0J_0J_0\rangle$ and $\langle J_2J_0J_0J_0\rangle$, where the scalar operator $J_0$ has twist $2$ and the operator $J_2$ is the stress-tensor.  From this example we show the four-point functions can be uniquely fixed by the approximate higher spin Ward identity. The approximate higher spin Ward identity can be separated into two equations according to their charges of parity symmetry.
The two equations relate the correlation functions at different orders of the 't Hooft coupling $\lambda$, therefore they provide recursion relations for the $\lambda$-expansions of these correlation functions. In fact the recursion relation terminates at order $\lambda^2$. This leads to solutions to the bootstrap equation exact in $\lambda$.
The solution of $\langle J_0J_0J_0J_0\rangle$ agrees with that obtained in \cite{Turiaci:2018dht}, while the solution of $\langle J_2J_0J_0J_0\rangle$ is new.
To solve the bootstrap equation, one of the major challenges is the rather complicated tensor structures arising from the spinning operators and the partial derivatives, although the tensor structures have already been notably simplified in the light-cone coordinate \cite{Maldacena:2011jn, Maldacena:2012sf}. Moreover, there are ambiguities in the tensor structures, that certain tensor structures with different forms are actually the same due to algebraic identities.
To solve these problems, we adopt the {\it conformal frame} which maximally uses the  conformal symmetry and fixes $10$ out of $12$ components in the $3D$ coordinates of the four external operators \cite{Mack:1976pa, Osborn:1993cr, Kravchuk:2016qvl}. By doing so the tensor structures in the equation turns out to be tractable.

We address the $3D$ bosonization duality using the result of spinning correlator $\langle J_2J_0J_0J_0\rangle$ in quasi-fermion theory. This will provide a new test for the bosonization duality at spinning four-point function level.
In the bosonization duality web, the CS-fermion theory relates to the CS-critical boson theory. It interpolates from  free fermion theory to critical boson theory by tuning the 't Hooft coupling $\lambda$ from $ 0$ to $\infty$. The solution of the spinning four-point function $\langle J_2J_0J_0J_0\rangle$ in quasi-fermion theory can be schematically written as\footnote{In \cite{Turiaci:2018dht}, this form has been proposed as an ansatz for the spinning four-point function.}
\beq
\langle J_2J_0J_0J_0\rangle=\frac{1}{\sqrt{1+\lambda^2}}\langle J_2J_0J_0J_0\rangle^{\textrm{ff}}+\frac{\lambda}{\sqrt{1+\lambda^2}}\langle J_2J_0J_0J_0\rangle^{\textrm{odd}},
\eeq
where the $\langle\cdots\rangle^{\textrm{ff}}$ term is parity odd and given by the free fermion theory, while the $\langle\cdots\rangle^{\textrm{odd}}$ term only appears in the theories with slightly broken higher spin symmetry and is parity even.
Explicit solutions of the two parts are provided in Eq.\ref{j2even}/\ref{j24peven} and Eq.\ref{j2odd}/\ref{oddsolution}. In the large $\lambda$ limit only the odd part remains and according to the bosonization duality, this should give the $\langle J_2J_0J_0J_0\rangle$ correlation function in critical-boson theory. An independent computation of this correlator in critical-boson theory is not known according to our knowledge. But it relates to the scalar four-point function in the free boson theory through a Legendre transformation \cite{Gubser:2002vv}. We show that the result nicely agrees with the solution to the approximate higher spin Ward identity.
Therefore the odd part of the correlator $\langle J_2J_0J_0J_0\rangle$ in quasi-fermion theory, which arises due to the slightly breaking of higher spin symmetry, is totally determined by the free boson theory through the bosonization duality and Legendre transformation! The bosonization duality provides a straightforward approach to compute part of the four-point functions, which is significantly easier than solving the approximate higher spin Ward identity.

The paper is organized as follows. In section $2$ we briefly review the aspects of (slightly broken) higher spin symmetry and the (pseudo) charge conservation law. In section $3$ we introduce the ingredients in the bootstrap equation and assumptions used in the bootstrap approach, from which we solve the approximate higher spin Ward identity of scalar four-point correlator. In section $4$ we show that the same result can be obtained using $3D$ bosonization and Legendre transformation. We conclude the results and discuss future directions in section $5$. In our computations we need to use various types of conformal integrals and the properties of $\Db$-functions. Details on the conformal integrals are presented in the appendices.

\section{Higher spin algebra and Ward identity}
For completeness we briefly review the higher spin algebra and the higher spin Ward identity in conformal theories.
This approach has been developed in \cite{Maldacena:2011jn, Maldacena:2012sf} and is remarkably powerful in constraining $3D$ CFTs with (slightly broken) higher spin symmetry. Actually in the theories with exact higher spin symmetry, by imposing the conservation condition of higher spin currents $J_s$ ($s>2$), the three-point correlators can be restricted to compact forms  \cite{Giombi:2011rz}. Specifically, in \cite{Giombi:2011rz} it has been shown that the scalar operator $J_0$ has to have dimension $\Delta_0=1$ or $2$ in order to have non-vanishing three-point correlators
with conserved higher spin currents $J_s$, and the three-point correlators of conserved currents admit following structure
\bea
\langle J_{s_1}J_{s_2}J_{s_3}\rangle=\alpha~ \langle J_{s_1}J_{s_2}J_{s_3}\rangle_{\textrm{fb}}+\beta~\langle J_{s_1}J_{s_2}J_{s_3}\rangle_{\textrm{ff}}+\gamma~\langle J_{s_1}J_{s_2}J_{s_3}\rangle_{\textrm{odd}}, \label{thpt}
\eea
in which the subscript ``fb" (``ff") indicates the correlator generated by free boson (fermion) theory, while the ``odd" term is not generated by any free theories and is parity violating. The odd term is non-vanishing only for the spins satisfying the triangle rule
$
s_i\leqslant s_{i+1}+s_{i+2}
$.
These results provide strong evidence that the theories with exact higher spin symmetry are likely to be free. Nevertheless, the odd term remains mysterious, and to understand the role of this term, it needs to explore more dynamical restrictions from higher spin symmetry. This is what has been fulfilled in \cite{Maldacena:2011jn, Maldacena:2012sf}.

\subsection{Exact higher spin symmetry}
The analysis in \cite{Maldacena:2011jn} starts from general consistency conditions of $3D$ CFTs with exact higher spin symmetry, such as the operator product expansion, unitarity, existence of a stress tensor, etc, and the conclusion applies to general microscopic theories with higher spin symmetry. The crucial ingredients in the analysis are the higher spin algebra and the Ward identity corresponding to the higher spin symmetry. Assuming the higher spin symmetry is generated by a local current $J_{\mu_1\mu_2...\mu_s}$ with spin $s>2$, then one can construct the conserved charges in the following way: firstly by contracting $J_{\mu_1\mu_2...\mu_s}$ with spin $s-1$ conformal Killing tensor $\epsilon^{\mu_1\mu_2...\mu_{(s-1)}}$, it gives a spin $1$ conserved current $j_\mu=J_{\mu\mu_1\mu_2...\mu_{(s-1)}} \epsilon^{\mu_1\mu_2...\mu_{(s-1)}}$ and the higher spin charge $Q_s$ is given by
$Q_s=\int_{\Sigma}*j$,
where $\Sigma$ is the codimension 1 hypersurface and $*j$ is the dual of the current $j_\mu$. Due to the conservation of current $j_\mu$, the charge $Q_s$ does not change by shifting $\Sigma$ without passing any extra operators. The conformal $n$-point correlators satisfy following charge conservation identity, or the higher spin Ward identity
\beq
0=Q_s\langle \cO_1(x_1)\cO_2(x_2)...\cO_n(x_n) \rangle=\sum_i\langle \cO_1(x_1)...[Q_s,\cO_i(x_i)]...\cO_n(x_n) \rangle.
\label{hswi}
\eeq
This is the main bootstrap equation for CFTs with higher spin symmetry. In this work, we always assume the external operators locate at different positions therefore we do not need to worry about the contact terms.

Following \cite{Maldacena:2011jn}, we choose the light cone coordinates in $3D$ spacetime as $ds^2=dx^+dx^-+dy^2$.  It turns out that for many problems, it is sufficient to consider just the $x_-$ component in the tensor structures. This amounts to pick up a special conformal killing tensor $\epsilon$ whose non-vanishing component is $\epsilon^{-...-}$.
Moreover, in \cite{Maldacena:2011jn} it shows that in any theory with a conserved higher spin current $j_s$ ($s>2$), there is always a conserved spin $4$ current, therefore the $Q_4$ charge conservation identity is ubiquitous in any theories with higher spin symmetry. Following the conventions discussed above, the $Q_4$ charge is given by
\bea
Q_4=\int_{x_-=\text{const}}dx^-dy J_{----},
\eea
which has spin $3$ and twist $0$.

Before we can obtain anything concrete from (\ref{hswi}), we need to know how the higher spin charge $Q_4$ acts on each operator. The action of charge $Q_4$ on the current $J_s$ with twist $\tau\equiv\Delta-\ell=1$ can be written as
\bea
[Q_4,J_s]=\sum_{s^\prime} c_{s,s^\prime} \partial_-^{s+3-s^\prime}J_{s^\prime},  \label{Q4j}
\eea
where $J_{s^\prime}$s are also twist $1$ conserved currents, as required by twist conservation. The range of spin $s^\prime$ is determined by the conservation of spin and the associativity of current algebra. Actually not all the currents appear in the right hand side (RHS) of (\ref{Q4j}) are consistent with the charge conservation equations. This corresponds to $c_{s,s^\prime}=0$ for these currents.
The coefficients $c_{s,s^\prime}$ are further fixed by imposing the charge conservation equations of the three-point correlators among the conserved currents.

From (\ref{hswi}) and (\ref{Q4j}), we can see that by demanding the n-point correlator is annihilated by $Q_4$, it gives a conservation equation which relates correlators of operators with different spins. Then we can impose charge conservation on these correlators respectively, which leads to a set of new equations, etc. Finally by using the charge conservation, we obtain a large set of equations for the $n$-point correlation functions. In \cite{Maldacena:2011jn} the authors show that these equations actually restrict the whole theory to be free, constructed from either free bosons or free fermions depending on the twist of scalar current $J_0$.

\subsection{Slightly broken higher spin symmetry}
Remarkably, the higher spin Ward identity also provides strong constraint on the theories with slightly broken higher spin symmetry \cite{Maldacena:2012sf}. With broken higher spin symmetry, the higher spin currents are not conserved
\bea
\partial\cdot J_s=g~ \cO_{s-1}. \label{divs}
\eea
The divergence $\cO_{s-1}$ is a conformal primary with twist $3$ in the limit $g\rightarrow0$. From the conformal representation point of view, the conserved current $J_s$ belongs to a shortened conformal multiplet; by turning on the interaction, this conformal multiplet becomes a long multiplet by recombining with another multiplet $\cO_{s-1}$.
 A crucial assumption associated with the scenario of slightly  symmetry breaking is that there is no twist $3$ single trace operators in the theory, in another word, the divergence $\cO_{s-1}$ does not contain single trace operators. The current operator $J_s$ is Higgsed by double trace operators,
and it acquires anomalous dimensions at order $1/N$ through the quantum effects. The general form of $\cO_{s-1}$ is given by the double or triple trace operators with twist $3$ and the relative coefficients among these terms are fixed in order to give a conformal primary operator. In this work, we will use the non-conservation equation of the current $J_4$ in quasi-fermion theory
\bea
\partial_\mu J^\mu_{---}=g\, (\partial_- J_0 J_2-\frac{2}{5}J_0 \partial_- J_2), \label{div4}
\eea
where the quasi-fermion scalar current $J_0$ has twist $2$ and $g\propto 1/N$ with normalization $\langle J_s J_s\rangle\sim N$ \cite{Maldacena:2012sf}. The three-point functions of these non-conserved currents still have the compact form (\ref{thpt}), while the odd terms are modified that they could be nonzero even outside of the triangle $s_i\leqslant s_{i+1}+s_{i+2}$.

The non-conserved higher spin currents lead to a modified version of the charge conservation identity (\ref{hswi}), which can be derived by inserting the divergence of $J_4$ (\ref{divs}) in an $n$-point correlator
\bea
\langle (\nabla\cdot J_s(x)) \cO_1\cO_2...\cO_n\rangle=g \langle  \cO_{s-1}(x)\cO_1\cO_2...\cO_n\rangle. \label{noncon}
\eea
The non-conservation identity of correlator $\langle \cO_1\cO_2...\cO_n\rangle$ is obtained from above equation by integrating over variable $x$. The integral diverges at the points where $x$ coincides with the coordinates $x_i$ of external operators $\cO_i$. The divergences can be regularized by introducing spherical boundaries of the integration domain $S_i$ which encloses $x_i$, and then taking the radius $r_i\rightarrow0$.
The RHS of (\ref{noncon}) involves double trace operators and can be factorized in the planar limit. The integral over $x$ on the left side of (\ref{noncon}) only gives boundary terms
\bea
\sum_{i=1}^{n}\langle  \int_{S_i}\boldsymbol{e}^\mu J_{\mu---}(x) \cO_1..\cO_i..\cO_n\rangle,
\eea
where $\boldsymbol{e}$ is the normal vector of $S_i$. The finite part of the integral over each sphere $S_i$ is equivalent to the action of non-conservation charge $Q_4$ on the operator $\cO_i$: $[Q_4,\cO_i]$. The approximate higher spin Ward identity takes the form:
\bea
\sum_{i=1}^n \langle \cO_1...[Q_4,\cO_i]...\cO_n\rangle=g \int d^3x \langle\cO_{s-1}(x)\cO_1\cO_2...\cO_n\rangle. \label{nhswi}
\eea
This is an integro-differential equation. In \cite{Maldacena:2012sf} the identities for three-point correlators have been solved which lead to a complete classification of all the three-point functions. In particular the three coefficients in (\ref{thpt})
are fixed to be\footnote{In \cite{Skvortsov:2018uru} the same result is obtained by an alternative bootstrap approach, which solves the planar three-point functions by constructing a non-linear realization of the conformal algebra.}
\beq
\alpha=\frac{\lambda^2}{1+\lambda^2}, ~~~~~\beta=\frac{1}{1+\lambda^2}, ~~~~~
\gamma=\frac{\lambda}{1+\lambda^2}.
\eeq
It is expected the equation (\ref{nhswi}) can also be used to solve the four-point functions in the planar limit. In this work, we initiate this project by solving the approximate higher spin Ward identity of scalar four-point correlator $\langle J_0J_0J_0J_0\rangle$ in quasi-fermion theory.

\section{Four-point functions from approximate higher spin Ward identity}
In this section we solve the approximate higher spin Ward identity of the correlator $\langle J_0J_0J_0J_0\rangle$ in quasi-fermion theory. By applying the non-conserved higher spin charge $Q_4$ on the correlator, there is another spinning correlator $\langle J_2J_0J_0J_0\rangle$  appearing in the approximate higher spin Ward identity and both of them can be solved from the equation.

\subsection{Approximate higher spin Ward identity for the scalar correlator}
We aim to solve the simplest case of the approximate higher spin Ward identity (\ref{nhswi}), in which all the external operators are scalars $J_0$.
According to the results in \cite{Maldacena:2012sf}, the action of $Q_4$ on the scalar current $J_0$ in quasi-fermion theory is
\bea
[Q_4,J_0]=\partial^3_-J_0+\frac{1}{\sqrt{1+\lambda^2}}\epsilon_{-\mu\nu}\partial_-\partial^\mu J^\nu_-.
\eea
Here we follow the normalization of the scalar operator $J_0$ in \cite{GurAri:2012is, Turiaci:2018dht}: $\langle J_0(x_1) J_0(x_2)\rangle=1/x_{12}^{4}$, which differs from the normalization in \cite{Maldacena:2012sf} by a rescaling $$J_0\rightarrow \sqrt{\frac{1+\lambda^2}{N}}J_0.$$
Note that in this normalization, the $1/N$ factor in the divergence of the current non-conservation equation (\ref{div4}) is absorbed by the operators while the parameter $g$ is of order $O(N^0)$ now.
As will be shown later, it is convenient to solve the bootstrap equation and study the bosonization duality in this normalization. The constant $g$ in (\ref{div4}) is solved from approximate higher spin Ward identity of three-point correlator \cite{Maldacena:2012sf}
\bea
g=g_0 \frac{\lambda}{\sqrt{1+\lambda^2}},
\eea
where $g_0$ is a numerical coefficient  and independent of $\lambda$ and $N$.
Combining all the ingredients together, the approximate higher spin Ward identity (\ref{nhswi}) turns into following specific form
\bea
\partial^3_-\langle J_0(x_1)J_0(x_2)J_0(x_3)J_0(x_4) \rangle &+& \frac{1}{\sqrt{1+\lambda^2}}\epsilon_{-\mu\nu}\partial_-\partial^\mu\langle J_-^{\nu}(x_1)J_0(x_2)J_0(x_3)J_0(x_4) \rangle \nn\\
&& \hspace{-2cm}+ (1\leftrightarrow2)+(1\leftrightarrow3) +(1\leftrightarrow4) ~ \nn\\
&& \hspace{-5cm} = g_0\frac{\lambda}{\sqrt{1+\lambda^2}}  \int d^3x  \langle(\partial_- J_0 J_2-\frac{2}{5}J_0 \partial_- J_2)(x)J_0(x_1)J_0(x_2)J_0(x_3)J_0(x_4)\rangle \nn\\
&& \hspace{-5cm} =\frac{7}{5}g_0\frac{\lambda}{\sqrt{1+\lambda^2}}  \int d^3x  \langle\partial_- J_0(x)J_0(x_1) \rangle \langle J_2(x)J_0(x_2)J_0(x_3)J_0(x_4)\rangle \nn\\
&& \hspace{-2cm}+ (1\leftrightarrow2)+(1\leftrightarrow3) +(1\leftrightarrow4).  \label{qfhswi}
\eea
This is the bootstrap equation to solve the four-point functions.
In the last step we introduce the factorized expansion of the $5$-point correlator in the third line. Here the planar limit is assumed implicitly. Extra possible factorizations have no contributions on the integral. For instance, the three-point correlator $\langle J_0(x)J_0(x_1)J_0(x_2)\rangle$ vanishes. Such three-point correlator could have a contact term \cite{GurAri:2012is}, while it does not appear here for separated external points $x_1\neq x_2$. Another factorization with three-point correlator $\langle(\partial_- J_0 J_2-\frac{2}{5}J_0 \partial_- J_2)(x)J_0(x_1)J_0(x_2)\rangle$ also vanishes, as the double trace operator $\partial_- J_0 J_2-\frac{2}{5}J_0 \partial_- J_2$ is a spin $3$ conformal primary to leading order in $1/N$.

\subsection{General form of spinning four-point correlator}
The approximate higher spin Ward identity (\ref{qfhswi}) involves the spinning four-point correlator $\langle J_2(x_1)J_0(x_2)J_0(x_3)J_0(x_4) \rangle$.
In this four-point function, there are two linearly independent tensor structures with opposite parity  \cite{Maldacena:2011jn}
\bea
\langle J_2(x_1)J_0(x_2)J_0(x_3)J_0(x_4)\rangle&=& \nn\\
&&\hspace{-2.4cm} \langle J_2(x_1)J_0(x_2)J_0(x_3)J_0(x_4)\rangle^{\textrm{even}}+\langle J_2(x_1)J_0(x_2)J_0(x_3)J_0(x_4)\rangle^{\textrm{odd}},~~
\eea
where the ``even" term includes an $\epsilon-$tensor and is parity odd  while the ``odd" term is parity invariant. These tensor structures can be simply written down in an index-free notation in embedding space \cite{Costa:2011mg}.

Conformal symmetry of CFT in $3D$ Euclidean space $SO(4,1)$ can be linearly realized in the embedding space $\mathbb{M}^{5}$. We use $X_\mu$  to denote the coordinate in $\mathbb{M}^{5}$, and the scalar product of two variables is
\beq
X_{ij}\equiv -2X_i^\mu \cdot X_{j\mu}=x_{ij}^2.
\eeq
In the spinning four-point function, we will use the basic three-point tensor structure $Q_{i,jk}$
\beq
Q_{i,jk}=\frac{(Z_i\cdot X_j)(X_i\cdot X_k)-(Z_i\cdot X_k)(X_i\cdot X_j)}{X_j\cdot X_k},
\eeq
where $Z_i$ is the auxiliary vector from which one can reproduce the tensor indices of variable $X_i$.
In the embedding space $\mathbb{M}^{5}$, we can construct a parity odd scalar using the $SO(4,1)$ invariant $\epsilon-$tensor and $5$ vectors ($Z_1, X_i$) that appears in the four-point correlator $\langle J_2(X_1,Z_1)J_0(X_2)J_0(X_3)J_0(X_4) \rangle$:
\beq
\epsilon(Z_1,X_1,X_2,X_3,X_4)=\epsilon_{\mu\nu\rho\sigma\delta} Z_1^\mu X_1^\nu X_2^\rho X_3^\sigma X_4^\delta. \label{oddts}
\eeq
In $3D$ spacetime above parity odd tensor structure reduces to the form proportional to $\epsilon_{\mu\nu\rho}Q_{1,23}^\nu Q_{1,24}^\rho$.
Not all the three-point tensor structures $Q_{i,jk}$ are linearly independent. In our case,
 the correlator $\langle J_2(X_1,Z_1)J_0(X_2)J_0(X_3)J_0(X_4) \rangle$ only includes two linearly independent $Q_{1,ij}$ \cite{Costa:2011mg}. Product of two parity odd tensor structures is equivalent to multiple three-point tensor structures, as suggested by the identity
\beq
\epsilon_{\mu_1...\mu_5}\epsilon^{\nu_1...\nu_5}=5!~\delta_{[\mu_1}^{\nu_1}\cdots\delta_{\mu_5]}^{\nu_5}.
\eeq
This linear dependence can also be shown clearly by expanding the parity odd tensor structure (\ref{oddts}) in terms of $3D$ $\epsilon-$tensor and vectors.

The general forms associated with the two tensor structures in embedding space are
\bea
\langle
J_2(x_1)J_0(x_2)J_0(x_3)J_0(x_4)\rangle^{\textrm{even}}&=&
\frac{\epsilon(Z_1,X_1,X_2,X_3,X_4)}{X_{13}^3X_{12}X_{14}X_{24}^2} \nn\\
&& (Q_{1,23}\,g(u,v)+Q_{1,34}\,g(v,u)+Q_{1,42}\frac{1}{u^2}g(\frac{v}{u},\frac{1}{u})),~~~~~\label{j2even}
\\\nn\\
\langle
J_2(x_1)J_0(x_2)J_0(x_3)J_0(x_4)\rangle^{\textrm{odd}}&=&
\frac{1}{X_{13}^2X_{12}^{3/2}X_{14}^{3/2}X_{24}^{1/2}} \nn\\
&&(Q_{1,23}^2 \,h(u,v)+Q_{1,34}^2 \,h(v,u)+Q_{1,42}^2 \frac{1}{u^{\frac{1}{2}}}h(\frac{v}{u},\frac{1}{v})),  \label{j2odd}
\eea
where the cross ratios $u,v$ are given by
\beq
u=\frac{x_{12}^2x_{34}^2}{x_{13}^2x_{24}^2},~~~~~~~~~~~~v=\frac{x_{14}^2x_{23}^2}{x_{13}^2x_{24}^2}.
\eeq
Above formulas have exhausted all the possible linearly independent tensor structures. In (\ref{j2even}) we keep $3$ $Q_{1,ij}$s in the expression to show the permutation symmetry explicitly, but actually only two of them are linearly independent.

Due to the permutation symmetry among the three variables $x_{2/3/4}$, there is only one undetermined functions of cross ratios associated with each tensor structure in (\ref{j2even}) and (\ref{j2odd}). Moreover, these functions satisfy following conditions
\bea
g(u,v)&=&\frac{1}{u^{2}}g(\frac{1}{u},\frac{v}{u}), \\
h(u,v)&=&\frac{1}{u^{1/2}}h(\frac{1}{u},\frac{v}{u}). \label{constaint2}
\eea

\subsection{Solving the approximate higher spin Ward identity}
Now we reach the central part of this work: to solve the approximate higher spin Ward identity (\ref{qfhswi}).
This is an integro-differential equation. The four-point correlators which we aim to solve appear on both sides of the equation, as a result we actually do not have any concrete information on either side of the equation. A possible approach to get rid of this barrier is to transform the equation into momentum space, in which the integro-differential equation is simplified to an algebraic equation \cite{Turiaci:2018dht}. While as also commented in \cite{Turiaci:2018dht}, this approach suffers from problems about conformal invariance and the contact terms in momentum space. Alternatively,  the Mellin space seems to be a natural choice to study the planar correlators with slightly broken higher spin symmetry
\cite{Mack:2009mi, Penedones:2010ue, Fitzpatrick:2011ia, Paulos:2011ie}. One may think about to solve the bootstrap equation in Mellin space, where the integro-differential equation turns into a functional equation and it could be completely solved with certain ansatz on the possible forms of the solution. The Mellin space has been successfully applied to solve the correlation functions in \MyRoman{2}B supergravity on $AdS_5\times S^5$, see e.g. \cite{Rastelli:2016nze, Rastelli:2017udc} and the Polyakov bootstrap, see e.g. \cite{Gopakumar:2016wkt, Gopakumar:2016cpb}.
In our problem, the obstacle is that the correlation functions include polynomials of $u,v$, and their transformations in Mellin space are subtle. Besides, the tensor structures in the bootstrap equation (\ref{nhswi}) are difficult to control in the Mellin space.\footnote{For recent work on the spinning four-point functions in Mellin space, see \cite{Faller:2017hyt, Chen:2017xdz, Sleight:2018epi, Sleight:2018ryu}.}

In this work, we stay in coordinate space, and our strategy to solve the bootstrap equation (\ref{qfhswi}) is as follows:
\newline
\textbf{a}.  We show the equation (\ref{qfhswi}) is separated into two independent parts according to their charges under parity transformation $\mathbb{P}$.
\newline
\textbf{b}. We treat the approximate higher spin Ward identity (\ref{qfhswi}) as a perturbative equation. We expand the four-point functions in terms of the parameter $\lambda$.
The equation (\ref{qfhswi}) provides a recursion relation which connects the RHS conformal integral at order $O(\lambda^n)$ to the  left hand side (LHS) correlators at order $O(\lambda^{n+1})$. At each order $O(\lambda^n)$, the integro-differential equation is simplified into a nonhomogeneous differential equation.
We start the recursive procedure from correlators in free fermion theory and solve the equation order by order.
\newline
\textbf{c}. At each order $O(\lambda^n)$ of the equation (\ref{qfhswi}), we argue that the solution is unique up to a redundant term from free theory.

Let us discuss these steps in more detail.

In \textbf{a}, we claimed the approximate higher spin Ward identity (\ref{qfhswi}) is consisted of two independent parts. This can be seen from two ways. One is simply counting the tensor indices of each term in the equation (\ref{qfhswi}). We will give more details on this later in the explicit computations.
Here we provide an argument based on parity symmetry.
In the free fermion theory, the scalar operator $J_0$ is parity odd, and whole theory preserves parity symmetry.
In the theories with slightly broken higher spin symmetry, although the parity symmetry is broken by the interactions--the CS couplings, we can effectively restore the parity symmetry by assigning an odd parity to the 't Hooft coupling $\lambda$ (or the CS level $k$ in CS-vector models) \cite{Aharony:2011jz, Giombi:2011kc, Maldacena:2012sf}:
\beq
\mathbb{P}\,\lambda=-\lambda, ~~~~~~~~~\mathbb{P}\,k=-k.
\eeq
For this reason, the correlation functions associated with even tensor structures, which have the same parity as the correlator, such as $\langle J_0J_0J_0J_0\rangle$ and $\langle J_2J_0J_0J_0\rangle^\textrm{even}$ only obtain corrections at the orders $O(\lambda^n)$ with $n$ even, while the correlator
$\langle J_2J_0J_0J_0\rangle^\textrm{odd}$ only acquires corrections at the orders $O(\lambda^n)$ with $n$ odd.
Consequently, the approximate higher spin Ward identity is separated into two sets of equations at orders of $O(\lambda^{2n})$ or $O(\lambda^{2n+1})$, which can be schematically written as follows:
\bea
\partial^3_-\langle J_0J_0J_0J_0 \rangle &+& \frac{1}{\sqrt{1+\lambda^2}}\epsilon_{-\mu\nu}\partial_-\partial^\mu\langle J_-^{\nu}J_0J_0J_0\rangle^{\textrm{even}} +\cdots \nn\\
&&   =\frac{7}{5}g_0\frac{\lambda}{\sqrt{1+\lambda^2}}  \int d^3x  \langle\partial_- J_0(x)J_0 \rangle \langle J_2(x)J_0J_0J_0\rangle^{\textrm{odd}}+\cdots,  \label{eveneq} \\
\frac{1}{\sqrt{1+\lambda^2}}&\epsilon_{-\mu\nu}&\partial_-\partial^\mu\langle J_-^{\nu}J_0J_0J_0\rangle^{\textrm{odd}} +\cdots \nn\\
&&   =\frac{7}{5}g_0\frac{\lambda}{\sqrt{1+\lambda^2}}  \int d^3x  \langle\partial_- J_0(x)J_0 \rangle \langle J_2(x)J_0J_0J_0\rangle^{\textrm{even}}+\cdots, \label{oddeq}
\eea
where we have omitted extra terms with permutated variables.

As briefly explained in \textbf{b}, we solve the approximate higher spin Ward identity (\ref{qfhswi}) perturbatively with respect to the parameter $\lambda$. The perturbation solution is governed by two separated equations (\ref{eveneq}) and (\ref{oddeq}). The advantage of the perturbative equations is that the conformal integral in the RHS can be directly evaluated by inserting the spinning four-point function solved at lower order. Thus we simplify the integro-differential equation (\ref{qfhswi}) into two nonhomogeneous differential equations.

In the bootstrap equation (\ref{qfhswi}), we have a non-polynomial factor $1/\sqrt{1+\lambda^2}$. This factor persists in (\ref{eveneq}) while is cancelled in (\ref{oddeq}). We do not take $\lambda$-series expansion of this factor otherwise it would complexify the equation unnecessarily. At leading order in $\lambda$, the equation (\ref{eveneq}) could take the following form
 \beq
 \partial^3_-\langle J_0J_0J_0J_0\rangle_{\lambda^0} + \frac{1}{\sqrt{1+\lambda^2}}\epsilon_{-\mu\nu}\partial_-\partial^\mu\langle J_-^{\nu}J_0J_0J_0 \rangle_{\lambda^0}^{\textrm{even}}+(1\leftrightarrow 2,3,4)=0, \label{qfhswi0}
 \eeq
 where $\langle...\rangle_{\lambda^0}$ denotes the correlator at order $O(\lambda^0)$.
Above equation differs from the exact higher spin Ward identity by an overall factor $1/\sqrt{1+\lambda^2}$, which can be absorbed by the spinning four-point correlator $\langle J_2J_0J_0J_0\rangle^{\textrm{even}}$ and the leading order equation still gives solutions of free fermion theory, up to a factor $\sqrt{1+\lambda^2}$. However, here we have an ambiguity in choosing the coefficient of the spinning correlator in the equation (\ref{qfhswi0}). Actually the factors $1/\sqrt{1+\lambda^2}$ and $\sqrt{1+\lambda^2}$ only differ by a higher order term $\lambda^2/\sqrt{1+\lambda^2}$\,:
\bea
&&\left\{\partial^3_-\langle J_0J_0J_0J_0 \rangle + \frac{1}{\sqrt{1+\lambda^2}}\epsilon_{-\mu\nu}\partial_-\partial^\mu\langle J_-^{\nu}J_0J_0J_0\rangle\right\} \longleftrightarrow \\
&&\left\{\partial^3_-\langle J_0J_0J_0J_0 \rangle +\sqrt{1+\lambda^2} \, \epsilon_{-\mu\nu}\partial_-\partial^\mu\langle J_-^{\nu}J_0J_0J_0\rangle\right\}- \frac{\lambda^2}{\sqrt{1+\lambda^2}}\epsilon_{-\mu\nu}\partial_-\partial^\mu\langle J_-^{\nu}J_0J_0J_0\rangle,~~~~~ \label{mfeveneq}
\eea
where the forms in the braces are the two candidates for the bootstrap equation at leading order.
We should get the same final result by choosing either one of them.
The difference only appears in the intermediate steps. In the next section we will show that the second one is more convenient to solve the equation (\ref{eveneq}). The recursion relation started from (\ref{mfeveneq}) terminates at order $O(\lambda^2)$ and provides an exact solution to the bootstrap equation.

Note in (\ref{mfeveneq}), by modifying the coefficient to $\sqrt{1+\lambda^2}$, we generate a second term at order $O(\lambda^2)$. This term will play a crucial role in solving the bootstrap equation at order $O(\lambda^2)$.

The claim \textbf{c} is important for us to make the whole computations realizable. Assuming we have already solved the equation (\ref{qfhswi}) at order $O(\lambda^{n-1})$, then at the next order $O(\lambda^n)$, the  RHS is obtained from a conformal integral of four-point correlator $\langle J_2J_0J_0J_0\rangle$ at order $O(\lambda^{n-1})$. The equation (\ref{qfhswi}) at order $O(\lambda^n)$ turns into a nonhomogeneous version of the Ward identity from exact higher spin symmetry. Given any two different solutions to this equation, by subtracting them, the nonhomogeneous term in the RHS cancels, and the difference should satisfy the higher spin Ward identity, which as proved in \cite{Maldacena:2011jn}, has a unique solution generated from free fermion theory. The conclusion is that at each order of approximate higher spin Ward identity, there is a unique solution up to an extra term from free theory. Because of this statement, we can obtain the one and only non-trivial solution at each order by constructing the solution explicitly and show it satisfying the equation, instead of solving a third order partial differential equation with complicated tensor structures.

In \cite{Turiaci:2018dht} the authors proposed to solve the approximate higher spin Ward identity with a promising ansatz for the four-point function $\langle J_2J_0J_0J_0\rangle$. The ansatz is based on the crossing symmetry, the behavior in the Regge limit, and more importantly, the consistency condition that its OPE expansion should reproduce the known three-point CFT data. The major challenge is the putative contact terms which break the analyticity in spin. Here we try not to use assumptions on the contact terms, instead we prove that they are absent in the four-point functions using the approach explained above.

Now we start the recursive procedure to solve the approximate higher spin Ward identity (\ref{qfhswi}).

\subsubsection{Leading order of approximate higher spin Ward identity}
To leading order $O(\lambda^0)$, the RHS of (\ref{qfhswi}) vanishes, and the equation (\ref{qfhswi}) goes back to the case with exact higher spin symmetry up to an overall scale factor $\sqrt{1+\lambda^2}$ for the correlator $\langle J_2J_0J_0J_0\rangle$. As discussed before, the leading order of the approximate higher spin Ward identity is given by the first term in (\ref{mfeveneq}), and equation is as follows
\beq
\partial^3_-\langle J_0J_0J_0J_0\rangle_{\lambda^0}  + \sqrt{1+\lambda^2}\,\epsilon_{-\mu\nu}\partial_-\partial^\mu\langle J_-^{\nu}J_0J_0J_0 \rangle_{\lambda^0} +(1\leftrightarrow 2,3,4)=0.
\eeq
The equation has a unique solution given by the free fermion theory \cite{Maldacena:2011jn}. The scalar four-point correlator is
\bea
\langle J_0(x_1)J_0(x_2)J_0(x_3)J_0(x_4)\rangle_{\lambda^0} &=&\frac{1}{x_{12}^4x_{34}^4}f(u,v) ,\nn\\
f(u,v)&=&a(1+u^2+\frac{u^2}{v^2})+ b \frac{\sqrt{u}}{v^{3/2}}\times    \label{scalar4} \\
  &&(u^{5/2}+v^{5/2}-u^{3/2} (v+1)- v^{3/2}(u+1)-u-v+1). \nn
\eea
In $f(u,v)$, the term proportional to $a\sim O(1)$ is from the disconnected diagrams and the second term is from connected diagrams with $b\sim 1/N$. The spinning four-point correlator $\langle J_2(x_1)J_0(x_2)J_0(x_3)J_0(x_4) \rangle_{\lambda^0}$ is also solved from the same equation. Solution of this four-point function takes the general form given by (\ref{j2even}) and (\ref{j2odd}), and the $u,v$-dependent functions are
\bea
g(u,v)&=&\frac{9}{5}\,\frac{b}{\sqrt{1+\lambda^2}}\, \frac{v}{u^{3/2}},  \label{j24peven}\\
h(u,v)&=&0.  \label{j24podd}
\eea
Note the function $g(u,v)$ has been rescaled by a factor $1/\sqrt{1+\lambda^2}$ comparing with the free fermion theory.
Exact numerical value of the coefficients $a,b$ can be fixed from the normalization of $J_0$ and the OPE of the four-point functions. The odd term vanishes at this level. This is expected since the odd term has even parity while the four-point correlator $\langle J_2(x_1)J_0(x_2)J_0(x_3)J_0(x_4) \rangle$ is parity odd in the free fermion theory. The parity-breaking interaction appears at the subleading order $O(\lambda)$, like the CS coupling at level $k$ in the regular fermion theory. This leads to non-zero contributions on the odd term of four-point correlator $\langle J_2(x_1)J_0(x_2)J_0(x_3)J_0(x_4) \rangle$.

\subsubsection{Approximate higher spin Ward identity at order $O(\lambda)$}
At this order, the equation (\ref{qfhswi}) is a nonhomogeneous modification of the higher spin Ward identity. The nonhomogeneous term is given by the conformal integral in the RHS
\beq
\frac{\lambda}{\sqrt{1+\lambda^2}}\int d^3x  \langle\partial_- J_0(x)J_0(x_1) \rangle \langle J_2(x)J_0(x_2)J_0(x_3)J_0(x_4)\rangle^{\textrm{even}}_{\lambda^0}+ (1\leftrightarrow2,3,4).
\eeq
The four-point function $\langle J_2(x)J_0(x_2)J_0(x_3)J_0(x_4)\rangle^{\textrm{even}}_{\lambda^0}$ in the integrand is given by the leading order solution (\ref{j24peven}). Unfortunately, according to our knowledge, there is no covariant approach for conformal integrations with spinning conformal four-point function. The covariant approach based on embedding space has been shown to be quite useful for conformal correlators. We expect there could be a covariant approach for conformal integrations, from which above conformal integral can be evaluated in a compact way.  In \cite{Liu:2018jhs} a covariant approach for conformal integrations with spinning three-point correlators has been developed, one probably can generalize their work to conformal integrations with generic four-point functions. In this work, we carry out the conformal integrations term by term in position space.

Inserting the solutions (\ref{j24peven},\ref{j24podd}) in the conformal integration, we have
\bea
\int d^3x_0  \langle\partial_- J_0(x_0)J_0(x_1) \rangle \langle J_2(x_0)J_0(x_2)J_0(x_3)J_0(x_4)\rangle^{\textrm{even}}_{\lambda^0} &\simeq& \nn\\
&& \hspace{-8.7cm}\frac{1}{\sqrt{1+\lambda^2}}\int d^3x_0 \left(\partial_-\frac{1 }{x_{01}^4}\right)\epsilon _{-\nu \rho }\left(\frac{x_{02}^{\nu } x_{03}^{\rho }}{x_{02}^2 x_{03}^2}+\frac{x_{03}^{\nu } x_{04}^{\rho }}{x_{03}^2 x_{04}^2}+\frac{x_{04}^{\nu } x_{02}^{\rho }}{x_{04}^2x_{02}^2 }\right) \times \nn\\
&&\hspace{-5.7cm}
\left(\left(\frac{x_{02-}}{x_{02}^2}-\frac{x_{03-}}{x_{03}^2}\right)\frac{x_{04}^2 }{x_{02} x_{03} x_{24}^3 x_{34}^3}-(2\leftrightarrow4)-(3\leftrightarrow4)\right),~~~~ \label{confint}
\eea
where we have ignored the numerical coefficients like $b$.
We expand the tensor structures in the integrand and evaluate each integral independently. The computations can be reduced using permutation symmetry. The integration involves the star-triangle relations and properties of $\Db$-functions.
Details on these conformal integrals are provided in the appendices.

Now the problem is to solve a non-homogeneous differential equation. We need to find the unique non-trivial solution that corresponds to the specific non-homogeneous term. Firstly it is obvious that only odd part of the four-point correlator $\langle J_2(x_1)J_0(x_2)J_0(x_3)J_0(x_4)\rangle^{\textrm{odd}}$ is affected by this nonhomogeneous term. The reason is that each term in the conformal integral (\ref{confint}) has tensor structures consisted of four indices: $x_{ij-}^3 x_{ijy}$ ($4$-vectors). While in the LHS of the equation (\ref{qfhswi}), both the scalar four-point correlator and the even part of $\langle J_2(x_1)J_0(x_2)J_0(x_3)J_0(x_4)\rangle$ generate tensor structures with $3$-vectors.\footnote{At first sight the correlation function $\langle J_2(x_1)J_0(x_2)J_0(x_3)J_0(x_4)\rangle^\textrm{even}$ can also generate tensor structures consisted of five $(x_{ij})$s ($5$-vectors), however, these terms can be further reduced to the tensor structures with $3$-vectors. Similarly, in the LHS of (\ref{qfhswi}), the correlation function $\langle J_2(x_1)J_0(x_2)J_0(x_3)J_0(x_4)\rangle^\textrm{even}$ can generate tensor structures with $6$-vectors, while after taking the symmetries into account, all the $6$-vectors reduce to $4$-vectors.}
The same conclusion can be made by analyzing the parity of each tensor structures on both sides of the equation (\ref{qfhswi}).
Therefore the approximate higher spin Ward identity is simplified to the equation (\ref{oddeq}).
The general form of the correlation function $\langle J_2(x_1)J_0(x_2)J_0(x_3)J_0(x_4)\rangle^{\textrm{odd}}$ is given by (\ref{j2odd}) with only one unknown function $h(u,v)$. We make an ansatz on the putative solution for $h(u,v)$. The ansatz should satisfy the constraint from permutation symmetry (\ref{constaint2}). Then we fix the parameters in the ansatz from the equation (\ref{oddeq}). As argued before, this solution should be unique up to a term generated by free fermion theory, which vanishes for the odd part.

Although the approximate higher spin Ward identity is significantly simplified in the form (\ref{oddeq}), the remaining tensor structures are still too complex to be solved directly. The equation (\ref{oddeq}) is redundant as it shows explicit conformal covariance and also the permutation symmetry of four variables.
More subtle problems are from the ambiguities in the $4$-vectors: two different tensor structures  could  actually be equal with each other through certain algebraic transformations. Due to these ambiguities, it is difficult to fix the parameters in the ansatz of $h(u,v)$ by matching the ansatz with the nonhomogeneous term in the RHS of (\ref{oddeq}).

The redundancies in the equation (\ref{oddeq}) can be removed by taking the {\it conformal frame}, in which the conformal symmetry is gauge-fixed by choosing a special configuration. By choosing this frame,  we also fix the permutation symmetry and the ambiguities in the tensor structures. In this work, we use the {\it conformal frame} in which the coordinates $x_i$ of the four external operators locate at
\bea
x_1 &=& (0,0,0), \\
x_2 &=& (t,t,s), \\
x_3 &=& (1,1,0),  \\
x_4 &=& (L,L,0).
\eea
Here the variables $x_i$ are given in terms of light-cone components  $x^i=(x^+,x^-,x^y)$. The parameter $L$ will be sent to infinity. For a single correlator, this is done by taking following limit \cite{Kravchuk:2016qvl}
\beq
\lim_{L\rightarrow+\infty} L^{2 \Delta_4}\langle \cO(x_1)\cO(x_2)\cO(x_3)\cO(x_4)\rangle.
\eeq
This is equivalent to pick up the leading terms in the large $L$ limit, which are of order $L^{-2\Delta_4}$. In our case, the equation (\ref{qfhswi}) involves multiple correlators with different scaling dimensions of $\cO_4$, therefore there is no universal factor $L^{2 \Delta_4}$. We pick up the leading terms of $L$ of the whole equation, which is of order $L^{-4}$ and the correlators with larger $\Delta_4$ vanish in this limit.
The equation (\ref{oddeq}) only depends on the two variables $s,t$. Then we insert the ansatz on $h(u,v)$ in the approximate higher spin Ward identity (\ref{oddeq}).
The solution of the correlation function $\langle J_2(x_1)J_0(x_2)J_0(x_3)J_0(x_4)\rangle^{\textrm{odd}}_{\lambda}$ is given by the form (\ref{j2odd}) with function $h(u,v)$:
\beq
h(u,v)\sim\frac{\lambda}{\sqrt{1+\lambda^2}} \frac{\sqrt{v}}{u^{3/2}}(1+u-v)(1+u+3\,v).  \label{oddsolution}
\eeq
Here we have ignored the numerical factors like $g_0$.
This is the only correction on the two four-point correlators $\langle J_0J_0J_0J_0\rangle$ and $\langle J_2J_0J_0J_0\rangle$ at order $O(\lambda)$. Since at $\lambda=0$ there is no nonzero solution for the homogeneous version of equation (\ref{oddeq}), this solution is unique for general $\lambda\neq0$.

\subsubsection{Approximate higher spin Ward identity at order $O(\lambda^2)$}
In the last part we have obtained the correction to the four-point correlator $\langle J_2J_0J_0J_0\rangle^{\textrm{odd}}$ at the order $O(\lambda)$. This term contributes to the conformal integral in the RHS of (\ref{qfhswi}) at the order $O(\lambda)$
\beq
\frac{\lambda}{\sqrt{1+\lambda^2}}\int d^3x  \langle\partial_- J_0(x)J_0(x_1) \rangle \langle J_2(x)J_0(x_2)J_0(x_3)J_0(x_4)\rangle^\textrm{odd}_{\lambda}+ (1\leftrightarrow2,3,4).
\eeq
Consequently, it generates corrections on the four-point correlators in the LHS of (\ref{qfhswi}) at the order of $O(\lambda^2)$. In this case the conformal integral leads to tensor structures of the form $x_{ij-}^3$, which only appear in the correlators  $\langle J_0J_0J_0J_0\rangle$ and $\langle J_2J_0J_0J_0\rangle^{\textrm{even}}$ in the LHS of the equation (\ref{qfhswi}). Therefore at this order,  only corrections on the correlators $\langle J_0J_0J_0J_0\rangle$ and  $\langle J_2J_0J_0J_0\rangle^{\textrm{even}}$ are possible, as expected from the argument of restored parity symmetry.
Combining the forms in (\ref{eveneq}) and (\ref{mfeveneq}), the approximate higher spin Ward identity at this order can be written as
\begin{equation}
\begin{aligned}
\partial^3_-\langle J_0J_0J_0J_0 \rangle_{\lambda^2} &+ \sqrt{1+\lambda^2}\epsilon_{-\mu\nu}\partial_-\partial^\mu\langle J_-^{\nu}J_0J_0J_0\rangle^{\textrm{even}}_{\lambda^2}  \\ &-\frac{\lambda^2}{\sqrt{1+\lambda^2}}\epsilon_{-\mu\nu}\partial_-\partial^\mu\langle J_-^{\nu}J_0J_0J_0\rangle^{\textrm{even}}_{\lambda^0}\cdots \\
&\hspace{0cm}=\frac{7}{5}g_0\frac{\lambda}{\sqrt{1+\lambda^2}}  \int d^3x  \langle\partial_- J_0(x)J_0 \rangle \langle J_2(x)J_0J_0J_0\rangle^{\textrm{odd}}_{\lambda}+\cdots, \label{qfhswi2}
\end{aligned}
\end{equation}
in which the correlators $\langle \cdots\rangle_{\lambda^2}$ refer to the possible corrections at the order $\lambda^2$.

Again there is no covariant approach for the conformal integral at the RHS of (\ref{qfhswi2}) and we have to evaluate it term by term. The conformal integration involves some special types of four-point conformal integrals. The relevant identities for these integrations are provided in the appendices.
This integral does not give any new ingredients but just reproduce the correlator in the second line of (\ref{qfhswi2}): $-\lambda^2/\sqrt{1+\lambda^2}\epsilon_{-\mu\nu}\partial_-\partial^\mu\langle J_2J_0J_0J_0\rangle^{\textrm{even}}_{\lambda^0}$. Therefore at this order, there is no nontrivial corrections on the correlators $\langle J_0J_0J_0J_0\rangle$ and $\langle J_2J_0J_0J_0\rangle^{\textrm{even}}$. The recursion relation ends up at this order, and we obtain exact solution of the approximate higher spin Ward identity (\ref{qfhswi}).

\section{Bosonization duality at the level of four-point functions}
In this section, we use the four-point function $\langle J_2J_0J_0J_0\rangle$ to test the $3D$ bosonization duality in the planar limit.
Strong evidence on the large $N$ $3D$ bosonization duality has been given at the level of three-point correlators of single trace currents \cite{Maldacena:2012sf,Aharony:2012nh,GurAri:2012is}. This duality has also been tested using the scalar four-point correlator $\langle J_0J_0J_0J_0\rangle$ computed with Feynman diagrams in both quasi-fermion and quasi-boson theories \cite{Bedhotiya:2015uga, Turiaci:2018dht, Yacoby:2018yvy}.
The four-point correlators receive contributions from both single and double trace operators, from which we can compute dynamical data on the double trace operators, such as their anomalous dimension and OPE coefficients, thereby the four-point functions can provide highly nontrivial test on the bosonization duality.

\setlength{\unitlength}{1 mm}
\begin{figure}
\begin{center}
\begin{picture}(150,80)
\thicklines
\put(45,71){\vector(4,0){50}}
\put(45,71){\vector(-1,0){0}}
\put(98,31){\vector(0,4){20}}
\put(120,25){\vector(0,4){35}}
\put(10,70){Free fermion}
\put(110,70){Critical boson}
\put(110,15){Free boson}
\put(88,25){$\langle J_2J_0J_0J_0\rangle_{\text{fb}}^{\text{even}}$}
\put(88,55){$\langle J_2J_0J_0J_0\rangle_{\text{cb}}^{\text{odd}}$}
\put(97,62){$0$}
\put(28,55){$0$}
\put(18,62){$\langle J_2J_0J_0J_0\rangle_{\text{ff}}^{\text{even}}$}
\put(50,60){\vector(-1,0){0}}
\put(50,60){\vector(1,0){30}}
\put(123,40){RG flow}
\put(58,74){CS coupling}
\end{picture}
\vspace{-1.5cm}
\caption{A schematic description of the duality map among the free fermion (ff), critical boson (cb) and free boson (fb) theories. We also provide the duality transformations of the correlation functions associated with different tensor structures. In a more complete web of the bosonization duality, there is another corner for critical fermion theory, and it relates to solution to the quasi-boson approximate higher spin Ward identity, which is not studied in this work.}
 \label{diag1}
 \end{center}
\end{figure}
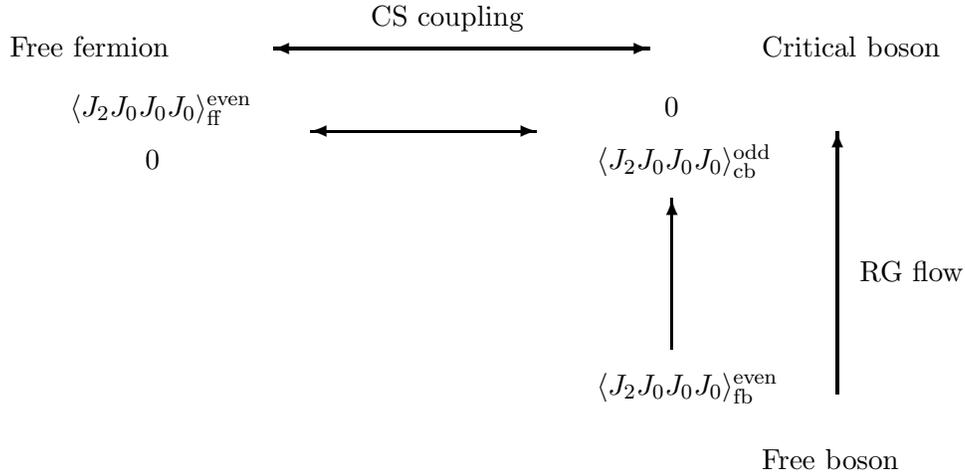

In the web of the $3D$ bosonization duality, the CS-regular fermion (boson) theory is related to CS-critical boson (fermion) theory. In our case, we work with three theories of free fermion, critical boson and free boson, which are connected with each other through CS coupling or RG flow. Their relations and the mapping among the spinning correlators are schematically explained in Fig \ref{diag1}. In the planar limit, the CS-regular fermion theory interpolates from the free fermion theory to the critical boson theory.
According to our result on the four-point correlator $\langle J_2J_0J_0J_0\rangle$, the even and odd terms are given by
\bea
\langle J_2J_0J_0J_0\rangle^{\textrm{even}} &=& \frac{1}{\sqrt{1+\lambda^2}}\langle J_2J_0J_0J_0\rangle^{\textrm{even}}_{\textrm{ff}}, \\
\langle J_2J_0J_0J_0\rangle^{\textrm{odd}} &=& \frac{\lambda}{\sqrt{1+\lambda^2}}
\frac{1}{X_{13}^2X_{12}^{3/2}X_{14}^{3/2}X_{24}^{1/2}} \nn\\
&&(Q_{1,23}^2 \,h(u,v)+Q_{1,34}^2 \,h(v,u)+Q_{1,42}^2 \frac{1}{u^{\frac{1}{2}}}h(\frac{v}{u},\frac{1}{v})), \label{bosniodd}
\eea
where the function $h(u,v)$ is given in (\ref{oddsolution}).\footnote{Here we have stripped the $\lambda$-dependent factor out of $h(u,v)$, to show its behavior in the large $\lambda$ limit. }
The bosonization duality predicts that the four-point function $\langle J_2J_0J_0J_0\rangle$ in critical boson theory is obtained by taking the limit $\lambda\rightarrow \infty$ of above forms. In particular, the even term vanishes in this limit.
This prediction can be explicitly checked by computing the four-point function $\langle J_2J_0J_0J_0\rangle$ in critical boson theory.

The critical boson theory is realized as an IR fixed point of UV free boson theory perturbed by the double trace  operator $\cO^2$, where $\cO$ is the single trace scalar operator with scaling dimension $\Delta_\cO=1$. It has been known in the early stage of AdS/CFT that these UV/IR fixed points are corresponding with each other through a Legendre transformation. Holographically it corresponds to exchanging the two unitary boundary conditions of the bulk operator dual to $\cO$ \cite{Witten:2001ua, Gubser:2002vv}. The conformal correlation functions in critical boson theory can be easily computed based on the results in free boson theory by employing this UV/IR correspondence \cite{Leonhardt:2003du, Hartman:2006dy,Giombi:2011ya,Giombi:2017mxl,Giombi:2018vtc}.

\subsection{Scalar four-point function from Legendre transformation}
The scalar four-point correlator $\langle J_0J_0J_0J_0\rangle$ has been computed both in quasi-fermion and quasi-boson theories \cite{Bedhotiya:2015uga, Turiaci:2018dht, Yacoby:2018yvy}. The results are well consistent with the bosonization duality. Here we take this correlator as an example to show how the Legendre transformation  works for conformal four-point correlators. In \cite{Giombi:2017mxl,Giombi:2018vtc} there are similar examples for three-point correlators and four-point correlators in which the four external operators are different from $J_0$.

The Legendre transformation can be explicitly realized with the Hubbard-Stratonovich auxiliary field method. The action for IR CFT is
\beq
S_{IR}=S_{UV}+\int d^3x J_0\cO-\frac{1}{4\alpha}J_0^2,
\eeq
in which the action $S_{UV}$ is of free boson theory and the operator $J_0$ is an auxiliary field. By integrating out $J_0$, we reproduce the double trace deformation term $\alpha\,\cO^2$. In the IR limit the quadratic term of $J_0$ can be dropped, and the auxiliary field $J_0$ has an induced two-point function.
To leading order in $1/N$, this two-point function is
\beq
\langle J_0(x_1)J_0(x_2)\rangle\sim \frac{1}{x_{12}^4}. \label{J0prp}
\eeq
This is just the propagator of a conformal primary operator with scaling dimension $2$.

\setlength{\unitlength}{1 mm}

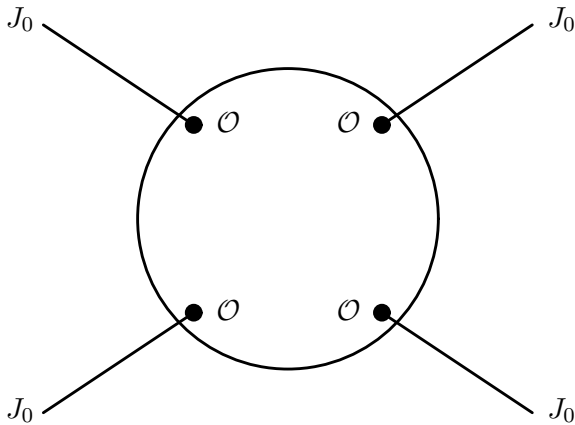
\begin{figure}
\begin{picture}(150,70)
\Thicklines
\put(60,45){\line(-3,2){20}}
\put(60,20){\line(-3,-2){20}}
\put(85,45){\line(3,2){20}}
\put(85,20){\line(3,-2){20}}
\linethickness{0.4mm}
\put(72.5,32.5){\circle{40}}
\put(60,45){\circle*{2}}
\put(60,20){\circle*{2}}
\put(85,45){\circle*{2}}
\put(85,20){\circle*{2}}
\put(35,58){$J_0$}
\put(35,6){$J_0$}
\put(107,58){$J_0$}
\put(107,6){$J_0$}
\put(63,44){$\cO$}
\put(79,44){$\cO$}
\put(63,19){$\cO$}
\put(79,19){$\cO$}
\end{picture}
\caption{Conformal four-point function $\langle J_0J_0J_0J_0\rangle$ in the critical boson theory obtained from Legendre transformation of the four-point function $\langle \cO \cO\cO\cO\rangle$ in the free boson theory. The enclosed part of the diagram represents the connected term of the correlator $\langle \cO \cO\cO\cO\rangle$. There are also diagrams with disconnected contractions of $\cO$s. The disconnected diagrams lead to the disconnected terms of $\langle J_0J_0J_0J_0\rangle$.
Here we are only interested in the connected part of the diagram.  } \label{fig4scalar}
\end{figure}

The conformal four-point correlator $\langle J_0J_0J_0J_0\rangle$ can be evaluated by contracting with four vertices $\int d^3x J_0\cO$.
Here we only pay attention to the connected part of the correlator, see Fig. \ref{fig4scalar}.
The diagram, up to an overall constant, takes the following form
\bea
\langle J_0(x_1)J_0(x_2)J_0(x_3)J_0(x_4)\rangle &=& \int d^3x_5d^3x_6d^3x_7d^3x_8 \times \nn\\
&& \langle J_0(x_1)J_0(x_2)J_0(x_3)J_0(x_4) J_0\cO(x_5)\cdots J_0\cO(x_8)\rangle. \label{Feydgm1}
\eea
In the planar limit, above eight-point correlator can be factorized into
\bea
\langle J_0(x_1)J_0(x_2)J_0(x_3)J_0(x_4) J_0\cO(x_5)\cdots J_0\cO(x_8)\rangle &\rightarrow& \nn\\
&&\hspace{-7cm}\langle  J_0(x_1)J_0(x_5)\rangle\cdots  \langle  J_0(x_4)J_0(x_8)\rangle \langle \cO(x_5) \cO(x_6) \cO(x_7) \cO(x_8)\rangle+{\textrm{discon...}},~~
\eea
in which we have omitted the contributions from disconnected diagrams.
Permutations of the $\langle J_0J_0\rangle$ contractions give the same results due to the permutation symmetry of the four-point correlator $\langle \cO(x_5) \cO(x_6) \cO(x_7) \cO(x_8)\rangle$.
Note that we do not have factorizations with three-point correlator like $\langle J_0(x_1)J_0(x_2)J_0(x)\rangle$, as such correlator (with two separated external variables) vanishes in quasi-fermion theories to leading order in $1/N$.
This is the reason why we need at least four vortices $\int d^3x J_0\cO$ in (\ref{Feydgm1}).
Using the induced propagator of $J_0$ (\ref{J0prp}), we obtain the following conformal integral
\beq
\int d^3x_5d^3x_6d^3x_7d^3x_8 \frac{1}{x_{15}^4}\frac{1}{x_{26}^4}\frac{1}{x_{37}^4}\frac{1}{x_{48}^4}
\langle\cO(x_5)\cO(x_6)\cO(x_7)\cO(x_8)\rangle.  \label{Feydgm2}
\eeq
This gives the Legendre transformation of the scalar four-point correlator.
The connected part of the conformal four-point function $\langle \cO\cO\cO\cO\rangle$ in the free boson theory is
\beq
\langle \cO(x_1)\cO(x_2)\cO(x_3)\cO(x_4)\rangle\sim\frac{1}{x_{13}^2x_{24}^2}(\frac{1}{\sqrt{u}}+
\frac{1}{\sqrt{v}}+\frac{1}{\sqrt{u\,v}}).
\eeq
Plugging this form in (\ref{Feydgm2}) and completing the conformal integration, we reproduce the result of the four-point correlator $\langle J_0J_0J_0J_0\rangle$ in the free fermion theory (\ref{scalar4}).

\subsection{Spinning four-point function from Legendre transformation}
In the free boson theory, the four-point correlator $\langle J_2\cO\cO\cO\rangle$ has been solved in \cite{Maldacena:2011jn}, and it takes the following form
\bea
\langle J_2(x_1)\cO(x_2)\cO(x_3)\cO(x_4)\rangle &=& \frac{x_{24}^2}{x_{13}^2x_{12}^4x_{14}^4} \nn\\
&&\times
(Q_{1,23}^2 g_*(u,v)+Q_{1,34}^2 g_*(v,u)+Q_{1,42}^2, u \,g_*(\frac{v}{u},\frac{1}{u})),  \label{freej2boson}\\
g_*(u,v)&\sim& \frac{9}{20}\frac{v^2}{\sqrt{u}}. \nn
\eea
The parity odd term vanishes in the free boson theory. Similar to the scalar four-point function, we can construct the four-point function $\langle J_2J_0J_0J_0\rangle$ from Legendre transformation of the result of free boson theory, and compare it with the solution to the approximate higher spin Ward identity.

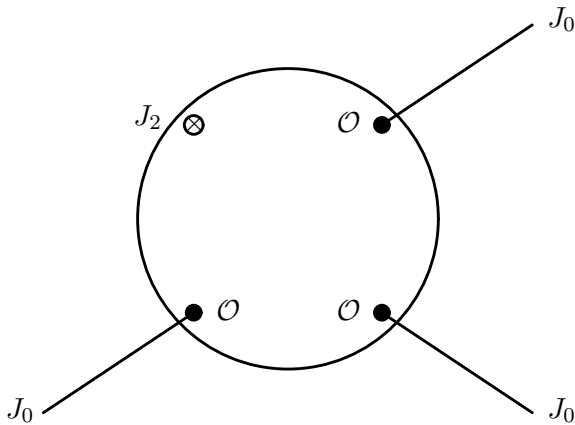
\begin{figure}
\begin{picture}(150,70)
\Thicklines
\put(60,20){\line(-3,-2){20}}
\put(85,45){\line(3,2){20}}
\put(85,20){\line(3,-2){20}}
\linethickness{0.4mm}
\put(72.5,32.5){\circle{40}}
\put(60,45){\circle{2.5}}
\put(58.6,44.2){$\times$}
\put(60,20){\circle*{2}}
\put(85,45){\circle*{2}}
\put(85,20){\circle*{2}}
\put(52,45){$J_2$}
\put(35,6){$J_0$}
\put(107,58){$J_0$}
\put(107,6){$J_0$}
\put(79,44){$\cO$}
\put(63,19){$\cO$}
\put(79,19){$\cO$}
\end{picture}
\caption{Conformal four-point function $\langle J_2J_0J_0J_0\rangle$ in critical boson theory obtained from Legendre transformation of the four-point function $\langle J_2 \cO\cO\cO\rangle$. The operator $J_2$ inside the circle is actually an external operator instead of from the vertex.} \label{diag2}
\end{figure}

In the correlator $\langle J_2J_0J_0J_0\rangle$, one of the external operators $J_2$ is the stress tensor. It appears in both  $UV$ and $IR$ theories and is not affected by the Legendre transformation. In this case we need to introduce three vortices $\int d^3x J_0\cO$ in the Legendre transformation. This has been shown diagrammatically in Fig. \ref{diag2}, which corresponds to the following form
\bea
\langle J_2(x_1)J_0(x_2)J_0(x_3)J_0(x_4)\rangle &=& \int d^3x_5d^3x_6d^3x_7 \times \nn\\
&& \langle J_2(x_1)J_0(x_2)J_0(x_3)J_0(x_4) J_0\cO(x_5)\cdots J_0\cO(x_7)\rangle \nn  \\
&& \hspace{-2cm}=\int d^3x_5d^3x_6d^3x_7 \frac{1}{x_{25}^4}\frac{1}{x_{36}^4}\frac{1}{x_{47}^4} \langle J_2(x_1)\cO(x_5)\cO(x_6)\cO(x_7)\rangle.
\label{Feydgm3}
\eea
In the last step we have applied the factorization of the seven-point correlator in the planar limit
\bea
\langle J_2(x_1)J_0(x_2)J_0(x_3)J_0(x_4) J_0\cO(x_5)\cdots J_0\cO(x_7)\rangle &\rightarrow& \nn\\
&&\hspace{-7cm}\langle  J_0(x_2)J_0(x_5)\rangle\cdots  \langle  J_0(x_4)J_0(x_7)\rangle \langle J_2(x_1)\cO(x_5) \cO(x_6) \cO(x_7)\rangle+\textrm{discon...}~~
\eea
We ignore the permutations of the $\langle J_0J_0\rangle $ contractions as they lead to the same result due to permutation symmetry of the correlator $ \langle J_2(x_1)\cO(x_5) \cO(x_6) \cO(x_7)\rangle$. Then plugging the four-point function (\ref{freej2boson}) in the conformal integral (\ref{Feydgm3}), and completing the conformal integration, we obtain the planar four-point function $\langle J_2J_0J_0J_0\rangle$ in the critical boson theory, which exactly matches the result (\ref{bosniodd}). The conformal integral shares similar forms as in the approximate higher spin Ward identity.   More details on the conformal integration are provided in the appendices.

Our computations in this part show that the correlator $\langle J_2J_0J_0J_0\rangle$ in critical boson theory shares the same form as in the quasi-fermion theory with 't Hooft coupling $\lambda\rightarrow \infty$. This is nicely consistent with the predictions from bosonization duality. A more complete test of this duality is to compute the four-point correlator in quasi boson theory with general 't Hooft coupling $\lambda$, and show that they correspond to the quasi-fermion correlator through Legendre transformation. This should be fulfilled by solving the approximate higher spin Ward identity for the quasi-boson theory, following the same procedure we used in this work for the quasi-fermion theories.

The results obtained here also provide a nice ``explanation" for the odd term of the four-point function $\langle J_2J_0J_0J_0\rangle$. This is a new term associated with slightly broken higher spin symmetry. Previously we solved this function from approximate higher spin Ward identity through rather cumbersome computations, while the final result is concise. One may  expect there is a simple reason hidden in the complicated computations in charge of this compact result. Now this has been clarified from bosonization duality: the odd term just arises from the Legendre transformation of the correlator in free boson theory! We expect this scenario also works for general planar single trace four-point correlators. This would uncover deep connection between the slightly broken higher spin symmetry and the bosonization duality.
A thorough study along this direction requires more detailed analysis on the tensor structures of the four-point functions and their  conformal integrations.

\section{Conclusions and discussions}
We have solved the planar four-point functions $\langle J_0J_0J_0J_0\rangle$ and $\langle J_2J_0J_0J_0\rangle$ in quasi-fermion theories using approximate higher spin Ward identity.
Our results support the speculation that the slightly broken higher spin symmetry can be used to uniquely fix the conformal $n$-point functions ($n\geqslant4$) in the planar limit \cite{Maldacena:2012sf}. This speculation is motivated by the remarkable success in  classification of the three-point functions using constraints from slightly broken higher spin symmetry \cite{Maldacena:2012sf}. For the four-point functions, the approximate higher spin Ward identity is given by a complicated integro-differential equation. We showed that this equation admits
 recursion relations for the perturbative expansions of the correlation functions $\langle J_0J_0J_0J_0\rangle$ and $\langle J_2J_0J_0J_0\rangle$. At leading order in $\lambda$, the solution is given by free fermion theory. A new term proportional to the odd tensor structure appears at the subleading order. The recursion relations terminate at the second order and we obtain solutions of the correlators $\langle J_0J_0J_0J_0\rangle$ and $\langle J_2J_0J_0J_0\rangle$ exact in 't Hooft coupling in the planar limit. We adopted the conformal frame to fix conformal gauge redundancies and ambiguities in the approximate higher spin Ward identity. The scalar four-point function $\langle J_0J_0J_0J_0\rangle$  agrees with the result obtained from Feynman diagrams in CS-fermion theory \cite{Bedhotiya:2015uga, Turiaci:2018dht}. The solution of the spinning correlator $\langle J_2J_0J_0J_0\rangle$ also proves an ansatz proposed in \cite{Turiaci:2018dht}.

We tested the $3D$ bosonization duality using the spinning four-point correlator $\langle J_2J_0J_0J_0\rangle$. We showed that the solution to the approximate higher spin Ward identity of quasi-fermion theory, in the large 't Hooft coupling limit, matches the planar spinning four-point function in the critical boson theory. The four-point functions in the critical boson theory were computed using Legendre transformation of free boson theory. This reproduces results from approximate higher spin Ward identity in a much simper and straightforward way.

The method used in this work is a typical bootstrap approach, in the sense that we only employed the algebra of slightly broken higher spin symmetry and consistency conditions of general unitary CFTs, while did not use any microscopic realizations of the theories. The bootstrap approach has been shown to be surprisingly powerful in solving higher dimensional CFTs \cite{Rattazzi:2008pe}, see \cite{Poland:2018epd} for a review. Comparing to computing specific CFT data, less work has been done towards solving four-point functions.
In \cite{Turiaci:2018dht} the authors attempted to apply the analytical bootstrap technique to fix the scalar four-point function. It turns out that the analyticity in spin plays an important role on this purpose.
In this work, we showed that contact terms that can break the maximal analyticity in spin do not appear in the spinning correlator $\langle J_2J_0J_0J_0\rangle$ as well. This agrees with the expectation in \cite{Turiaci:2018dht}, where
the authors argued that from the bulk locality point of view, the correlators of theories with slightly broken higher spin symmetry should admit the maximal analyticity in spin.

It would be interesting to interpret our bootstrap results in  higher spin gravity theories in $AdS_4$ and the CS-vector models. From the $AdS/CFT$ point of view, the four-point functions computed in this work correspond to the four-point amplitudes of Vasiliev theory in $AdS_4$, in which the higher spin symmetry is broken by the boundary conditions. It would be interesting to understand how the corresponding bulk interactions, such as the graviton and $3$ scalars scattering amplitude can be fixed in higher spin gravity theory. The connection has been partially revealed from the equivalence between the diagrams used in the  Legendre transformation for the $IR$ theories and the Witten diagrams in $AdS$ \cite{Hartman:2006dy, Giombi:2011ya, Giombi:2018vtc}.
In the quantum field theory side, our results is equivalent to compute the correlators of gauge invariant operators in the CS-vector models.
In both cases, the  explicit field theory realizations of slightly broken higher spin symmetry involve gauge interactions, and the computations are entangled with problem of gauge fixing. In contrast, the method used in this work solves the gauge invariant correlation functions in a perturbatively on-shell way and is free from the problem of gauge fixing.

There is an intriguing interplay between the bootstrap approach and the bosonization duality. The bootstrap approach, although has already been simplified comparing with the Feynman diagrams, still requires massive computations.
It is quite remarkable that the same results can be straightforwardly obtained using bosonization duality.
In both approaches, the critical ingredients for the final results are from the same type of conformal integrations. While in the approximate higher spin Ward identity, there are gauge redundancies that make the computations physically more obscure. Also it is of third-order differential equation, which is above the ``physical" differential orders of the equation of motions in field theory.
It is likely that the approximate higher spin Ward identity is endowed with a hidden structure that is more close to the essence of bosonization duality.

This work is the beginning of a more ambitious project on classification of all the planar four-point functions of single trace currents in CFTs with slightly broken higher spin symmetry. This object is reachable based on the bootstrap approach. It is straightforward to write down the approximate higher spin Ward identity for general spinning four-point correlators, based on the results in \cite{Maldacena:2012sf}, and in principle we should be able to solve these equations following the perturbative steps used in this work. The major challenges are from the tensor structures and their conformal integrations. Previous experiences and techniques developed for numerical bootstrap of spinning correlators could be helpful on this purpose. Also a covariant approach for the conformal  integrations of spinning four-point correlators can greatly promote this bootstrap project. Generically the tensor structures of these spinning four-point functions could be formidably complicated. While in the CFTs with slightly broken higher spin symmetry, we expect these tensor structures are well organized as mild modifications of the cases with exact higher spin symmetry, in which the correlation functions are given by invariants of bulk higher spin symmetry \cite{Didenko:2012tv}.
On the other hand, one could use bosonization duality or maximal analyticity in spin to figure out possible solutions of the spinning correlators.

By solving the general spinning four-point functions, we expect to obtain better understanding on the deep connections among the slightly broken higher spin symmetry, bosonization duality and the higher spin interactions in $AdS_4$. The knowledge of general spinning four-point functions
 $\langle J_{s_1}J_{s_2}J_{s_3}J_{s_4}\rangle$ is also needed to obtain CFT data at the order $1/N^2$ through analytic bootstrap, as these correlators contain information of mixing of the double trace operators $[J_{m}J_{n}]$ at order $1/N$ \cite{Aharony:2018npf}. The $3D$ large N Chern Simons theories have been shown to saturate the average null energy condition and certain stress tensor correlators, like $\langle J_2 J_2 J_0J_0\rangle$ are expected to be essentially free \cite{Chowdhury:2017vel, Cordova:2017zej}. It would be interesting to check this by solving these correlators explicitly.
A particularly interesting example is the stress-tensor four-point correlator $\langle J_2J_2J_2J_2\rangle$. This correlator appears in any theories with local interactions. It has been computed in \cite{Raju:2012zs} using bulk perturbation theory in momentum space. In \cite{Dymarsky:2017yzx} this correlator has been studied using modern numerical bootstrap. By imposing the consistency conditions from unitarity and crossing symmetry, the authors obtained strong constraints on the CFT data for general unitary conformal theories. In \cite{Karateev:2018oml} the authors studied the OPE coefficients of the stress-tensor correlator in Mean Field Theory.
The stress-tensor correlators play a central role in the conformal collider physics  \cite{Hofman:2008ar}.
Most of the work on conformal collider physics aimed to obtain bounds on the OPE coefficients from general consistency conditions. While in the scenario with slightly broken higher spin symmetry, the constraint is much stronger from which the stress-tensor four-point correlator could be completely solved, in consequence, it could unveil more concrete information on the bulk higher spin gravity theory.

\section*{Acknowledgements}
I would like to thank António Antunes, Miguel Costa, Shailesh Lal, Joao Penedones, Aaditya Salgarkar, Sourav Sarkar and Joao Silva  for useful discussions. I am also grateful to Sasha Zhiboedov for helpful communications.  This work was supported by the Simons Foundation grant 488637 (Simons collaboration on the Nonperturbative bootstrap). This research received funding from the grant CERN/FIS-PAR/0019/2017. Centro de F\'isica do Porto is partially funded by the Foundation for Science and Technology
of Portugal (FCT).


\appendix
\section{Conformal integration and $\Db$-function}
Our computations heavily rely on the conformal integrals. For completeness we briefly explain the definitions and conventions on the conformal integrals adopted in this work. We mainly use the notations of Dolan and Osborn \cite{Dolan:2000uw, Dolan:2000ut}.

We are interested in the following $n$-point conformal integral
\beq
\int d^dx \prod_{i=1}^n\frac{1}{(x-x_i)^{2\Delta_i}},~~~~~~~~~~~~~~\sum_{i} \Delta_i=d.
\eeq
For $n=3$, we have the well-known star-triangle relation
\bea
\int d^dx\frac{1}{(x-x_1)^{2\Delta_1}(x-x_2)^{2\Delta_2}(x-x_3)^{2\Delta_3}} &=& \nn\\
&&\hspace{-5cm}\pi^h\frac{\Gamma (h-\De_1) \Gamma (h-\De_2) \Gamma (h-\De_3)}{\Gamma (\De_1) \Gamma (\De_2) \Gamma (\De_3)} \frac{1}{x_{12}^{2(h-\De_3)}x_{13}^{2(h-\De_2)}x_{23}^{2(h-\De_1)}}, \label{star-tri}
\eea
where $h\equiv d/2$. More general three-point conformal integrals with spinning operators can be obtained by decomposing into scalar integrands with derivatives.

For $n=4$, the four-point conformal integral is
\begin{equation}
\begin{aligned}
\int d^d x &\frac{1}{(x-x_1)^{2\Delta_1}(x-x_2)^{2\Delta_2}(x-x_3)^{2\Delta_3}(x-x_4)^{2\Delta_4}}\\
&=\frac{\pi^{h}}{\Gamma(\Delta_1)\Gamma(\Delta_2)\Gamma(\Delta_3)\Gamma(\Delta_4)}
\frac{x_{14}^{2(h-\Delta_1-\Delta_4)} x_{34}^{2(h-\Delta_3-\Delta_4)}}{x_{13}^{2(h-\Delta_4)} x_{24}^{2\Delta_2}}
\bar{D}_{\Delta_1\Delta_2\Delta_3\Delta_4}(u,v).
\label{Db}
\end{aligned}
\end{equation}
The function $\bar{D}$ in the third line is the reduced version of the so-called $D$-function. The $\bar{D}$-function only depends on the conformal invariant cross ratios. In practical computations, it would be very useful to expand the function $\bar{D}(u,v)$ near $u=0, v=1$. This is fulfilled through the $H$ function
\beq
\bar{D}_{\Delta_1\Delta_2\Delta_3\Delta_4} (u,v) = H\left( \Delta_2, h -\Delta_4, \Delta_1+\Delta_2 - h +1 , \Delta_1+\Delta_2; u,v \right),
\eeq
and the $H$ function is defined as follow:
\begin{equation}
\begin{aligned}
H&(\alpha,\beta,\gamma,\delta;u.v)
=\frac{\Gamma(1-\gamma)}{\Gamma(\delta)} \Gamma(\alpha) \Gamma(\beta)
 \Gamma(\delta - \alpha) \Gamma(\delta - \beta) G(\alpha, \beta,\gamma,\delta;u , 1 -v) \\
&+ \frac{ \Gamma(\gamma-1)}{ \Gamma(\delta-2\gamma+2)} \Gamma(\alpha-\gamma+1)
 \Gamma(\beta-\gamma+1)  \Gamma(\delta-\gamma+\alpha+1) \Gamma(\delta-\gamma-\beta+1) \\
&\hspace{3cm}\times u^{1-\gamma} G(\alpha - \gamma+1, \beta- \gamma+1,2 - \gamma, \delta - 2\gamma+2;u, 1-v).
\end{aligned}
\end{equation}
in which the $G$-function has the series expansion near $u=0,v=1$:
\beq
G(\alpha,\beta,\gamma,\delta;u,1-v) = \sum_{m,n=0}^\infty \frac{(\delta-\alpha)_m (\delta-\beta)_m (\alpha)_{m+n} (\beta)_{m+n}}{ m! n!(\gamma)_m (\delta)_{2m+n} } u^m (1-v)^n.
\eeq

From the integral definition of $\bar{D}$-functions, it is easy to prove the following symmetry identities
\begin{equation}
\begin{aligned}
\Db_{\De_1 \De_2 \De_3 \De_4}(u,v) &= v^{\De_1+\De_4-\Sigma}\Db_{\De_2\De_1\De_4\De_3}(u,v) \\
&= u^{\De_3+\De_4-\Sigma}\Db_{\De_4\De_3\De_2\De_1}(u,v) \\
& = \Db_{\De_3 \De_2 \De_1 \De_4}(v,u)\\
&= \Db_{\Sigma - \De_3 \Sigma-\De_4 \Sigma - \De_1 \Sigma-\De_2} (u,v),
\end{aligned} \label{Dbsym}
\end{equation}
where $\Sigma=\sum_i\De_i/2$.  Note above identities work for general $\De_i$s even if $h\neq\Sigma$.
The $\Db$-functions also have the relation
\begin{equation}
\begin{aligned}
\De_4\Db_{\De_1 \De_2 \De_3 \De_4}(u,v) =& \Db_{\De_1 \De_2 \De_3+1\, \De_4+1}(u,v)+\Db_{\De_1 \De_2+1\, \De_3 \De_4+1}(u,v) \\
&+\Db_{\De_1+1\, \De_2 \De_3 \De_4+1}(u,v). \label{Db4}
\end{aligned}
\end{equation}
When one of the parameter $\De_i=0$, the four-point conformal integral degenerates to the three-point conformal integral. This amounts to the following limitation of the $\Db$-function \cite{Dolan:2004iy}:
\beq
\lim_{\De_4\rightarrow0}\Db_{\De_1 \De_2 \De_3 \De_4}(u,v)\De_4=
\Gamma(\Sigma-\De_1)\Gamma(\Sigma-\De_2)\Gamma(\Sigma-\De_3)u^{\De_3-\Sigma}v^{\De_3-\Sigma}. \label{Db3}
\eeq
For more analytical properties of the $\Db$-function, see \cite{Dolan:2000uw, Dolan:2000ut,Dolan:2004iy}.

\section{Conformal integrals in approximate higher spin Ward identity}
The conformal integrals play crucial roles in the approximate higher spin Ward identity. These conformal integrals involve various tensor structures. We do not have a covariant approach to accomplish the integrations. Practically, we expand these spinning conformal integrals in terms of the scalar conformal integrals, and the tensor structures are replaced by derivatives with respect to the external variables. All the three-point conformal integrations can be done straightforwardly, using the star-triangle relation (\ref{star-tri}). In the following part, we focus on the four-point conformal integrals appearing in the quasi-fermion approximate higher spin Ward identity. These four-point conformal integrals are actually corresponding to truncated polynomials of $u,v$ instead of the stand $\Db$-functions.
Similar conformal integrals also appear in the Legendre transformation of spinning four-point correlators.

\textbf{Four-point conformal integral \MyRoman{1}:}

We compute the following four-point conformal integral
\beq
\int d^dx_0 \,\frac{x_{04}^2}{x_{01}^{2 \De_1} \,x_{0}^{2 \De_2} \,x_{03}^{2 \De_3}}, ~~~~~~~~~~{\De_1+\De_2+\De_3=d+1}. \label{int1}
\eeq
This four-point conformal integral includes a parameter $\De_4=-1$, which makes this integral reducible and the $\Db$-function degenerates to the type of three-point integrals. There are several approaches to evaluate this integral. For instance, one can expand $x_{04}^2=x_{02}^2+2x_{02}\cdot x_{04}+x_{24}^2$, and then replace the tensor indices by derivatives. Or alternatively, we can solve this integration using the definition of $\Db$-function in (\ref{Db}). We can introduce a regularization factor to $\De_4$: $\De_4=-1+\epsilon$, and then take the limit $\epsilon\rightarrow$ in the series expansion of $\Db$-function.
Here we provide a simple method which only employs the properties of $\Db$-functions presented before.

We regularize the conformal integral by setting $\De_4\rightarrow-1+\epsilon$ with spacetime dimension $d\rightarrow 3+\epsilon$. From the definition (\ref{Db}), above four-point integral turns into
\begin{equation}
\begin{aligned}
\int d^dx_0 \,\frac{x_{04}^2}{x_{01}^{2 \De_1} \,x_{0}^{2 \De_2} \,x_{03}^{2 \De_3}}=&\frac{\pi^{h}}{\Gamma(\Delta_1)\Gamma(\Delta_2)\Gamma(\Delta_3)\Gamma(-1+\epsilon)}\\
&\times
\frac{x_{14}^{2(h-\Delta_1+1-\epsilon)} x_{34}^{2(h-\Delta_3+1-\epsilon)}}{x_{13}^{2(h+1-\epsilon)} x_{24}^{2\Delta_2}}
\bar{D}_{\Delta_1\Delta_2\Delta_3(\epsilon-1)}(u,v).
\label{Db}
\end{aligned}
\end{equation}
Here we have a zero point from the factor $1/\Gamma(-1+\epsilon)$.
From the identity (\ref{Db4}), we have
\begin{equation}
\begin{aligned}
\Db_{\De_1 \De_2 \De_3\, (\epsilon-1)}(u,v) =& -\Db_{\De_1 \De_2 \De_3+1\, \epsilon}(u,v)-\Db_{\De_1 \De_2+1\, \De_3 \epsilon}(u,v) \\
&-\Db_{\De_1+1\, \De_2 \De_3 \,\epsilon}(u,v)+O(\epsilon).
\end{aligned}
\end{equation}
Here we only need the leading order term in $\epsilon$. In the next examples, we will pay attention to the subleading order terms. Now the $\Db$-functions reduce to the case in (\ref{Db3}). The leading term in the limit $\epsilon\rightarrow 0$ has a singularity proportional to $1/\epsilon$, which cancels the zero factor $1/\Gamma(-1+\epsilon)$. The final result is
\begin{equation}
\begin{aligned}
\int d^dx_0 \,\frac{x_{04}^2}{x_{01}^{2 \De_1} \,x_{02}^{2 \De_2} \,x_{03}^{2 \De_3}}=&\pi^{h}\frac{\Gamma(h+1-\Delta_1)\Gamma(h+1-\Delta_2)\Gamma(h+1-\Delta_3)}
{\Gamma(\Delta_1)\Gamma(\Delta_2)\Gamma(\Delta_3)}\\
&\hspace{-2cm}\times x_{12}^{2 (\De_3-h-1)} x_{13}^{2( \De_2-h-1)} x_{23}^{2 (\De_1-h-1)}\left(\frac{x_{14}^{2} x_{23}^{2}}{h-\De_1}+\frac{x_{24}^{2} x_{13}^{2}}{h-\De_2}+\frac{x_{34}^{2} x_{12}^{2}}{h-\De_3}\right),
\end{aligned}
\end{equation}
where $h=d/2$.

\textbf{Four-point conformal integral \MyRoman{2}:}
\newline
The most challenging parts in the conformal integrations are of the type
\beq
\int d^3x_0~\frac{x_{01-}^2}{x_{01}^2 x_{02}^2 x_{03}^2 x_{04}^4},
\eeq
or similarly
\beq
\int d^3x_0~\frac{x_{01-}x_{02-}}{x_{01}^2 x_{02}^2 x_{03}^2 x_{04}^4}.
\eeq
Here we use the light-cone coordinate $x_{-/+/y}$. These integrals with tensor indices can be transformed to the scalar four-point integrals

\bea
\int d^3x_0~\frac{x_{02-}^2}{x_{01}^4 x_{02}^2 x_{03}^2 x_{04}^2}&=&\frac{1}{\epsilon(1+\epsilon)}\partial_{2-}^2 \left.\int d^{3-\epsilon}x_0 \,\frac{x_{02}^{2(1+\epsilon)}}{x_{01}^4 \, x_{03}^2\, x_{04}^2} \right|_{\epsilon\rightarrow0}, \label{int3} \\
\int d^3x_0~\frac{x_{02-}x_{03-}}{x_{01}^4 x_{02}^2 x_{03}^2 x_{04}^2}&=&\frac{1}{\epsilon^2}\partial_{2-}\partial_{3-}\left.\int d^{3+2\epsilon}x_0  \label{int2} \,\frac{1}{x_{01}^{4}\,x_{02}^{2\epsilon} \, x_{03}^{2\epsilon}\, x_{04}^2} \right|_{\epsilon\rightarrow0}. \label{int4}
\eea
The integral in the RHS of (\ref{int3}) is close to the integral (\ref{int1}), nevertheless, there is an extra $1/\epsilon$ pole generated by the tensor indices, and the scalar four-point integral at order $O(\epsilon)$ can have non-trivial contributions on the whole term. The singularity in $\epsilon$ stays in the final expression of the integral, it is canceled by combining with extra terms in the whole covariant tensor structure. Therefore these poles are actually ``unreal".
We expect in a covariant approach of conformal integration, these poles will not appear in the computations.

Then scalar four-point integral in the RHS of (\ref{int3}) leads to a function $\Db_{2\,(-1-\epsilon)\,1\,1}$. We expand this $\Db$-function near $\epsilon=0$ to the order of $O(1)$. After some algebra we obtain
\begin{equation}
\begin{aligned}
\Db_{2\,(-1-\epsilon)\,1\,1}&|_{\epsilon\rightarrow0} =  \\
&\frac{1}{2 \epsilon }\pi ^{3/2}(1-u+v)-\frac{1}{4} \pi ^{3/2} ((\gamma +\log (4)) (u-v-1)+u+4) \\
&\hspace{-0cm}+\frac{1}{16} \pi ^{3/2} (v-1) \left(3 (v-1) \, _3F_2\left(1,1,\frac{5}{2};3,3;1-v\right)-4 u \, _3F_2\left(1,1,\frac{5}{2};2,4;1-v\right)\right)\\
&\hspace{-0cm}+\frac{1}{2} \pi ^{3/2} \left(\frac{u (v+1)}{\left(\sqrt{v}+1\right)^2}+(1-u+v) \log \left(1-\frac{u}{\left(\sqrt{v}+1\right)^2}\right)\right) \\
& \hspace{-0cm}+\pi ^{3/2} \left(\sqrt{u} \left(\sqrt{v}+1\right)+(1-u+v) \coth ^{-1}\left(\frac{\sqrt{v}+1}{\sqrt{u}}\right)\right)+O(\epsilon), \label{res1}
\end{aligned}
\end{equation}
where the parameter $\gamma$ is the Euler constant. The generalized hypergeometric functions $_3F_2$ appear in above formula probably could be simplified in terms of elementary functions.
The scalar four-point conformal integral in the RHS of (\ref{int4}) contains a function $\Db_{2\,\epsilon\,\epsilon\,1}$. It can be expanded near $\epsilon=0$ as follows:
\begin{equation}
\begin{aligned}
\Db_{2\,\epsilon\,\epsilon\,1}|_{\epsilon\rightarrow0} &= -\frac{\pi ^{3/2} \left(\sqrt{u}+1\right)}{\sqrt{u} \epsilon }+\pi ^{3/2} \log \left(\left(\sqrt{v}+1\right)^2-u\right) \\
&\hspace{-1cm}+\frac{\pi ^{3/2} }{\sqrt{u}}\left(\log \left(\left(\sqrt{v}+1\right)^2-u\right)+2 \left(\sqrt{u}+1\right) \coth ^{-1}\left(\frac{\sqrt{v}+1}{\sqrt{u}}\right)+\gamma  \sqrt{u}-2 \sqrt{v}+\gamma \right) \\
&+O(\epsilon),
\end{aligned}
\end{equation}
Based on the above $\epsilon$ expansions of the $\Db$-functions, we can evaluate the conformal integrations with spinning tensor structures. Despite of the complicated intermediate formulas, the final result of the spinning conformal integration is quite simple. For instance, we can prove following identity using (\ref{res1})
\begin{equation}
\begin{aligned}
\int d^3x_0~&\frac{x_{01-}^2}{x_{01}^2 x_{02}^2 x_{03}^2x_{04}^2}\left(\frac{x_{12}^2}{x_{02}^2}
+\frac{x_{13}^2}{x_{03}^2}+
\frac{x_{14}^2}{x_{04}^2}\right)= \frac{1}{x_{23}x_{24}x_{34}}\times\\
&\left(
\frac{x_{34}^2 }{x_{23}^{2} x_{24}^{2}}x_{12-}^2
-\frac{2 x_{12-} x_{13-}}{x_{23}^{2}}+\frac{x_{24}^2 }{x_{23}^{2} x_{34}^{2}}x_{13-}^2-\frac{2 x_{12-} x_{14-}}{x_{24}^{2}}-\frac{2 x_{13-} x_{14-}}{x_{34}^{2}}+\frac{x_{23}^2 }{x_{34}^{2} x_{24}^{2}}x_{14-}^2\right).
\end{aligned}
\end{equation}
We hope the properties of the spinning conformal integrals will be more transparently clarified in a covariant approach for conformal integrations.

\end{document}